# Package 'PEkit'

November 22, 2021

**Title** Partition Exchangeability Toolkit

**Version** 1.0.0.1000

**Description** Bayesian supervised predictive classifiers, hypothesis testing, and parametric estimation under Partition Exchangeability are implemented. The two classifiers presented are the marginal classifier (that assumes test data is i.i.d.) next to a more computationally costly but accurate simultaneous classifier (that finds a labelling for the entire test dataset at once based on simultaneous use of all the test data to predict each label). We also provide the Maximum Likelihood Estimation (MLE) of the only underlying parameter of the partition exchangeability generative model as well as hypothesis testing statistics for equality of this parameter with a single value, alternative, or multiple samples. We present functions to simulate the sequences from Ewens Sampling Formula as the realisation of the Poisson-Dirichlet distribution and their respective probabilities.

**License** MIT + file LICENSE

**Encoding** UTF-8

**RoxygenNote** 7.1.1.9001

**Suggests** testthat (>= 3.0.0)

**Config/testthat/edition** 3

**Imports** stats (>= 4.1.0)

**NeedsCompilation** no

**Author** Ville Kinnula [aut],
Jing Tang [ctb] (<https://orcid.org/0000-0001-7480-7710>),
Ali Amiryousefi [aut, cre] (<https://orcid.org/0000-0002-6317-3860>)

**Maintainer** Ali Amiryousefi <ali.amiryousefi@helsinki.fi>

**Repository** CRAN

**Date/Publication** 2021-11-22 08:50:02 UTC

# R topics documented:

---

abundance                          *Vector of frequencies of frequencies*

---

### Description

A function to calculate the abundance vector, or frequencies of frequencies of discrete or partly discrete data vector x. The abundance vector is used as input in the functions `dPD()`, `MLEp()`, and `LMTp()`.

### Usage

```
abundance(x)
```

### Arguments

x                     Data vector x.

### Details

This function is equivalent to `table(table(x))`.

### Value

This function returns a named vector with the frequencies of the frequencies in the data vector x. The function `base::table(x)` returns a contingency table with the frequencies in the input data vector x as values. The `names(table(x))` are the unique values in data vector x. In `abundance(x)`, the unique values in `table(x)` become the names of the values, while the values themselves are the frequencies of the frequencies of data vector x.

### Examples

```
set.seed(111)
x<-rpois(10,10)
## The frequency table of x:
print(table(x))
## The frequency table of the frequency table of x:
abundance(x)
```

---

| classifier.fit | *Fit the supervised classifier under partition exchangeability* |

---

#### Description

Fits the model according to training data x, where x is assumed to follow the Poisson-Dirichlet distribution, and discrete labels y.

#### Usage

```
classifier.fit(x, y)
```

#### Arguments

| | |
|---|---|
| x | data vector, or matrix with rows as data points and columns as features. |
| y | training data label vector of length equal to the amount of rows in x. |

#### Details

This function is used to learn the model parameters from the training data, and gather them into an object that is used by the classification algorithms `tMarLab()` and `tSimLab()`. The parameters it learns are the Maximum Likelihood Estimate of the $\psi$ of each feature within each class in the training data. It also records the frequencies of the data for each feature within each class as well. These are used in calculating the predictive probability of each test data being in each of the classes.

#### Value

Returns an object used as training data objects for the classification algorithms `tMarLab()` and `tSimLab()`.

If x is multidimensional, each list described below is returned for each dimension.

Returns a list of classwise lists, each with components:

`frequencies`: the frequencies of values in the class.

`psi`: the Maximum Likelihood estimate of $\psi$ for the class.

#### Examples

```
## Create training data x and its class labels y from Poisson-Dirichlet distributions
## with different psis:
set.seed(111)
x1<-rPD(5000,10)
x2<-rPD(5000,100)
x<-c(x1,x2)
y1<-rep("1", 5000)
y2<-rep("2", 5000)
y<-c(y1,y2)
fit<-classifier.fit(x,y)
```

```
## With multidimensional x:
set.seed(111)
x1<-cbind(rPD(5000,10),rPD(5000,50))
x2<-cbind(rPD(5000,100),rPD(5000,500))
x<-rbind(x1,x2)
y1<-rep("1", 5000)
y2<-rep("2", 5000)
y<-c(y1,y2)
fit<-classifier.fit(x,y)
```

---

dPD                                          *The Poisson-Dirichlet distribution*

---

### Description

Distribution function for the Poisson-Dirichlet distribution.

### Usage

```
dPD(abund, psi)
```

### Arguments

abund           An abundance vector.

psi             Dispersal parameter $\psi$. Accepted input values are positive real numbers, "a" for
                absolute value $\psi$=1 by default, or "r" for relative value $\psi = n$, where $n$ is the
                size of the input sample.

### Details

Given an abundance vector abunds, calculates the probability of a data vector x given by the
Poisson-Dirichlet distribution. The higher the dispersal parameter $\psi$, the higher the amount of
distinct observed species. In terms of the paintbox process, a high $\psi$ increases the size of the con-
tinuous part $p_0$ of the process, while a low $\psi$ will increase the size of the discrete parts $p_{\neq 0}$.

### Value

The probability of the Poisson-Dirichlet distribution for the input abundance vector, e.g. an ex-
changeable random partition, and a dispersal parameter $\psi$.

### Examples

```
## Get a random sample from the Poisson Dirichlet distribution, and
## find the probability of such a sample with psi=5:
set.seed(111)
s <- rPD(n=100,psi=5)
a=abundance(s)
dPD(a, psi=5)
```

---

is.PD                            *Test for the shape of the distribution*

---

### Description

This function performs a statistical test on the null hypothesis that a given sample's underlying distribution is the Poisson-Dirichlet distribution. It calculates a test statistic that is then used to gain a p-value from an empirical distribution of the statistic from simulated samples from a PD distribution.

### Usage

```
is.PD(x, rounds)
```

### Arguments

| | |
|---|---|
| x | A discrete data vector. |
| rounds | How many samples are simulated to obtain the empirical distribution. |

### Details

The calculated test statistic is

$$W = \sum_{i=1}^{n} n_i^2/n,$$

which is calculated from the sample. Here $n_i$ are the frequencies of each unique value in the sample. The MLE of $\psi$ is then estimated from the sample with the function `MLEp()`, and an amount of samples equal to the input parameter `rounds` are generated with that estimate of $\psi$ and sample size $n$. The test statistic $W$ is then calculated for each of the simulated samples. The original $W$ is then given a p-value based on what percentage of the simulated $W$ it exceeds.

### Value

A p-value.

**Examples**

```
##Test whether a typical sample follows PD:
x<-rPD(100,10)
is.PD(x, 100)

##Test whether a very atypical sample where frequencies of different values
## are similar:

x<-c(rep(1, 200), rep(2, 200), rep(3, 200), rep(4, 200), rep(5,200))
is.PD(x,50)
```

---

MLEp                           *Maximum Likelihood Estimate of $\psi$*

---

**Description**

Numerically searches for the MLE of $\psi$ given an abundance vector with a binary search algorithm.

**Usage**

```
MLEp(abund)
```

**Arguments**

abund            An abundance vector.

**Details**

Numerically searches for the MLE of $\psi$ as the root of equation

$$K = \sum_{i=1}^{n} \psi/(\psi + i - 1),$$

where $K$ is the observed number of different species in the sample. The right side of the equation is monotonically increasing when $\psi > 0$, so a binary search is used to find the root. An accepted $\psi$ sets value of the right side of the equation within R's smallest possible value of the actual value of $K$.

**Value**

The MLE of $\psi$.

**References**

W.J. Ewens, The sampling theory of selectively neutral alleles, Theoretical Population Biology, Volume 3, Issue 1, 1972, Pages 87-112, ISSN 0040-5809, <doi: 10.1016/00405809(72)900354>.

## Examples

```
##Find the MLE of psi of the vector (1,2,2).
##The frequencies of the frequencies of the data vector are given as input:
MLEp(abundance(c(1,2,2)))

##Find the MLE of psi of a sample from the Poisson-Dirichlet distribution:
set.seed(1000)
x<-rPD(n=10000, psi=100)
MLEp(abundance(x))
```

---

MLEp.bsci                    *Bootstrap confidence interval for the MLE of $\psi$*

---

## Description

A bootstrapped confidence interval for the Maximum Likelihood Estimate for $\psi$.

## Usage

```
MLEp.bsci(x, level = 0.95, rounds = 1000, frac = 0.8)
```

## Arguments

| | |
|---|---|
| x | A data vector. |
| level | Level of confidence interval as number between 0 and 1. |
| rounds | Number of bootstrap rounds. Default is 1000. |
| frac | Percentage of data x used for each bootstrap round. 0.8 by default with accepted values between 0 and 1. |

## Value

The MLE of $\psi$ as well as lower and upper bounds of the bootstrap confidence interval.

## Examples

```
## Find a 95% -confidence interval for the MLE of psi given a sample from the
## Poisson-Dirichlet distribution:
x<-rPD(n=10000, psi=100)
MLEp.bsci(x, 0.95, 100, 0.8)
```

---

mult.sample.test *Test for $\psi$ of multiple samples*

---

### Description

Likelihood ratio test for the hypotheses $H_0: \psi_1 = \psi_2 = ... = \psi_d$ and $H_1: \psi_1 \neq \psi_2 \neq ... \neq \psi_d$, where $\psi_1, \psi_2, ..., \psi_d$ are the dispersal parameters of the $d$ samples in the columns of the input data array x.

### Usage

```
mult.sample.test(x)
```

### Arguments

x                The data array to be tested. Each column of x is an independent sample.

### Details

Calculates the Likelihood Ratio Test statistic

$$-2log(L(\hat{\psi})/L(\hat{\psi}_1, \hat{\psi}_2, ..., \hat{\psi}_d)),$$

where L is the likelihood function of observing the $d$ input samples given a single $\psi$ in the numerator and $d$ different parameters $\psi_1, \psi_2, ..., \psi_d$ for each sample respectively in the denominator. According to the theory of Likelihood Ratio Tests, this statistic converges in distribution to a $\chi^2_{d-1}$-distribution when the null-hypothesis is true, where $d-1$ is the difference in the amount of parameters between the considered models. To calculate the statistic, the Maximum Likelihood Estimate for $\psi_1$, $\psi_2$, ..., $\psi_d$ of $H_1$ and the shared $\psi$ of $H_0$ are calculated.

### Value

Gives a vector with the Likelihood Ratio Test -statistic Lambda, as well as the p-value of the test p.

### References

Neyman, J., & Pearson, E. S. (1933). On the problem of the most efficient tests of statistical hypotheses. Philosophical Transactions of the Royal Society of London. Series A, Containing Papers of a Mathematical Or Physical Character, 231(694-706), 289-337. <doi: 10.1098/rsta.1933.0009>.

### Examples

```
##Create samples with different n and psi:
set.seed(111)
x<-rPD(1200, 15)
y<-c( rPD(1000, 20), rep(NA, 200) )
z<-c( rPD(800, 30), rep(NA, 400) )
samples<-cbind(cbind(x, y), z)
##Run test
mult.sample.test(samples)
```

| rPD | *Random sampling from the Poisson-Dirichlet Distribution* |

**Description**

rPD samples randomly from the PD distribution with a given $\psi$ by simulating the Hoppe urn model.

**Usage**

```
rPD(n, psi)
```

**Arguments**

| n | number of observations. |
| psi | dispersal parameter. |

**Details**

Samples random values with a given $\psi$ from the Poisson-Dirichlet distribution by simulating the Hoppe urn model.

**Value**

Returns a vector with a sample of size $n$ from the Hoppe urn model with parameter $\psi$.

**References**

Hoppe, F.M. The sampling theory of neutral alleles and an urn model in population genetics. J. Math. Biology 25, 123–159 (1987). <doi: 10.1007/BF00276386>.

W.J. Ewens, The sampling theory of selectively neutral alleles, Theoretical Population Biology, Volume 3, Issue 1, 1972, Pages 87-112, ISSN 0040-5809, <doi: 10.1016/00405809(72)900354>.

**Examples**

```
## Get random sample from the PD distribution with different psi,
## and estimate the psi of the samples:
s1<-rPD(1000, 10)
s2<- rPD(1000, 50)
print(c(MLEp(abundance(s1)), MLEp(abundance(s2))))
```

---

sample.test                    *Lagrange Multiplier Test for $\psi$*

---

**Description**

Performs the Lagrange Multiplier test for the equality of the dispersion parameter $\psi$ of a sample. The null hypothesis of the test is $H_0 : \psi = \psi_0$, where $\psi_0$ is given as input here.

**Usage**

```
sample.test(abund, psi = "a")
```

**Arguments**

abund          An abundance vector of a sample.

psi            Target positive number $\psi_0$ to be tested. Accepted values are "a" for absolute value 1, "r" for relative value $n$ (sample size), or any positive number.

**Details**

Calculates the Lagrange Multiplier test statistic

$$S = U(\psi_0)^2 / I(\psi_0),$$

where $U$ is the log-likelihood function of $\psi$ and $I$ is its Fisher information. The statistic $S$ follows $\chi^2$-distribution with 1 degree of freedom when the null hypothesis $H_0 : \psi = \psi_0$ is true.

**Value**

The statistic $S$ and a p-value of the two-sided test of the hypothesis.

**References**

Radhakrishna Rao, C, (1948), Large sample tests of statistical hypotheses concerning several parameters with applications to problems of estimation. Mathematical Proceedings of the Cambridge Philosophical Society, 44(1), 50-57. <doi: 10.1017/S0305004100023987>

**Examples**

```
## Test the psi of a sample from the Poisson-Dirichlet distribution:
set.seed(10000)
x<-rPD(1000, 10)
## Find the abundance of the data vector:
abund=abundance(x)
## Test for the psi that was used, as well as a higher and a lower one:
sample.test(abund, 10)
sample.test(abund, 15)
sample.test(abund, 5)
sample.test(abund)       #test for psi=1
sample.test(abund, "r")  #test for psi=n
```

| tMarLab | *Marginally predicted labels of the test data given training data classification.* |
|---|---|

## Description

Classifies the test data x based on the training data object. The test data is considered i.i.d., so each data point is classified one by one.

## Usage

```
tMarLab(training, x)
```

## Arguments

| training | A training data object from the function `classifier.fit()`. |
|---|---|
| x | Test data vector or matrix with rows as data points and columns as features. |

## Details

Independently assigns a class label for each test data point according to a *maximum a posteriori* rule. The predictive probability of data point $x_i$ arising from class $c$ assuming the training data of size $m_c$ in the class arises from a Poisson-Dirichlet($\hat{\psi}_c$) distribution is:

$$\hat{\psi}_c / (m_c + \hat{\psi}_c),$$

if no value equal to $x_i$ exists in the training data of class $c$, and

$$m_{ci} / (m_c + \hat{\psi}_c),$$

if there does, where $m_{ci}$ is the frequency of the value of $x_i$ in the training data.

## Value

A vector of predicted labels for test data x.

## Examples

```
## Create random samples x from Poisson-Dirichlet distributions with different
## psis, treating each sample as coming from a class of its own:
set.seed(111)
x1<-rPD(10500,10)
x2<-rPD(10500,1000)
test.ind1<-sample.int(10500,500) # Sample test datasets from the
test.ind2<-sample.int(10500,500) # original samples
x<-c(x1[-test.ind1],x2[-test.ind2])
## create training data labels:
y1<-rep("1", 10000)
y2<-rep("2", 10000)
y<-c(y1,y2)

## Test data t, with first half belonging to class "1", second have in "2":
t1<-x1[test.ind1]
t2<-x2[test.ind2]
t<-c(t1,t2)

fit<-classifier.fit(x,y)

## Run the classifier, which returns
tM<-tMarLab(fit, t)

##With multidimensional x:
set.seed(111)
x1<-cbind(rPD(5500,10),rPD(5500,50))
x2<-cbind(rPD(5500,100),rPD(5500,500))
test.ind1<-sample.int(5500,500)
test.ind2<-sample.int(5500,500)
x<-rbind(x1[-test.ind1,],x2[-test.ind2,])
y1<-rep("1", 5000)
y2<-rep("2", 5000)
y<-c(y1,y2)
fit<-classifier.fit(x,y)
t1<-x1[test.ind1,]
t2<-x2[test.ind2,]
t<-rbind(t1,t2)

tM<-tMarLab(fit, t)
```

---

tSimLab                            *Simultaneously predicted labels of the test data given the training data
                                    classification.*

---

## Description

Classifies the test data x based on the training data object. All of the test data is used simultaneously
to make the classification.

**Usage**

```
tSimLab(training, x)
```

**Arguments**

| | |
|---|---|
| training | A training data object from the function `classifier.fit()`. |
| x | Test data vector or matrix with rows as data points and columns as features. |

**Details**

The test data are first labeled with the marginal classifier. The simultaneous classifier then iterates over all test data, assigning each a label by finding the maximum predictive probability given the current classification structure of the test data as a whole. This is repeated until the classification structure converges after iterating over all data.

**Value**

A vector of predicted labels for test data x.

**References**

Amiryousefi A. Asymptotic supervised predictive classifiers under partition exchangeability. . 2021. https://arxiv.org/abs/2101.10950.

Corander, J., Cui, Y., Koski, T., and Siren, J.: Have I seen you before? Principles of Bayesian predictive classification revisited. Springer, Stat. Comput. 23, (2011), 59–73, (<doi: 10.1007/s1122201192917>).

**Examples**

```
## Create random samples x from Poisson-Dirichlet distributions with different
## psis, treating each sample as coming from a class of its own:
set.seed(111)
x1<-rPD(1050,10)
x2<-rPD(1050,1000)
test.ind1<-sample.int(1050,50) # Sample test datasets from the
test.ind2<-sample.int(1050,50) # original samples
x<-c(x1[-test.ind1],x2[-test.ind2])
## create training data labels:
y1<-rep("1", 1000)
y2<-rep("2", 1000)
y<-c(y1,y2)

## Test data t, with first half belonging to class "1", second have in "2":
t1<-x1[test.ind1]
t2<-x2[test.ind2]
t<-c(t1,t2)

fit<-classifier.fit(x,y)

## Run the classifier, which returns
```

```
tS<-tSimLab(fit, t)

##With multidimensional x:
set.seed(111)
x1<-cbind(rPD(500,1),rPD(500,5))
x2<-cbind(rPD(500,10),rPD(500,50))
test.ind1<-sample.int(500,50)
test.ind2<-sample.int(500,50)
x<-rbind(x1[-test.ind1,],x2[-test.ind2,])
y1<-rep("1", 450)
y2<-rep("2", 450)
y<-c(y1,y2)
fit<-classifier.fit(x,y)
t1<-x1[test.ind1,]
t2<-x2[test.ind2,]
t<-rbind(t1,t2)

tS<-tSimLab(fit, t)
```

---

two.sample.test                   *Two sample test for $\psi$*

---

### Description

Likelihood ratio test for the hypotheses $H_0 : \psi_1 = \psi_2$ and $H_1 : \psi_1 \neq \psi_2$, where $\psi_1$ and $\psi_2$ are the dispersal parameters of two input samples s1 and s2.

### Usage

```
two.sample.test(s1, s2)
```

### Arguments

s1, s2            The two data vectors to be tested.

### Details

Calculates the Likelihood Ratio Test statistic

$$-2log(L(\hat{\psi})/L(\hat{\psi}_1, \hat{\psi}_2)),$$

where L is the likelihood function of observing the two input samples given a single $\psi$ in the numerator and two different parameters $\psi_1$ and $\psi_2$ for each sample respectively in the denominator. According to the theory of Likelihood Ratio Tests, this statistic converges in distribution to a $\chi_d^2$-distribution under the null-hypothesis, where $d$ is the difference in the amount of parameters between the considered models, which is 1 here. To calculate the statistic, the Maximum Likelihood Estimate for $\psi_1$, $\psi_2$ of $H_1$ and the shared $\psi$ of $H_0$ are calculated.

**Value**

Gives a vector with the Likelihood Ratio Test -statistic `Lambda`, as well as the p-value of the test `p`.

**References**

Neyman, J., & Pearson, E. S. (1933). On the problem of the most efficient tests of statistical hypotheses. Philosophical Transactions of the Royal Society of London. Series A, Containing Papers of a Mathematical Or Physical Character, 231(694-706), 289-337. <doi: 10.1098/rsta.1933.0009>.

**Examples**

```
##Create samples with different n and psi:
set.seed(111)
x<-rPD(500, 15)
y<-rPD(1000, 20)
z<-rPD(800, 30)
##Run tests
two.sample.test(x,y)
two.sample.test(x,z)
two.sample.test(y,z)
```

# Index